\documentclass[12pt]{article}
\usepackage{amsmath,amssymb,amsthm,booktabs}

\usepackage{mathrsfs}
\usepackage{geometry}
\usepackage[OT1]{fontenc}
\usepackage[utf8]{inputenc}
\usepackage[colorlinks,citecolor=blue,urlcolor=blue]{hyperref}
\usepackage{txfonts}
\usepackage{indentfirst}
\usepackage{multirow}
\usepackage{float,subfig}

\usepackage{bm}
\usepackage{euscript}
\usepackage{graphicx}
\usepackage{multicol}
\usepackage[usenames,dvipsnames,svgnames,table]{xcolor}

\usepackage[numbers]{natbib}
\usepackage[ruled]{algorithm2e}

\numberwithin{equation}{section} \theoremstyle{plain}
\newtheorem{theorem}{Theorem}[section]

\newtheorem{definition}{Definition}[section]
\newtheorem{remark}{Remark}[section]

\makeatletter
\def\ps@pprintTitle{%
  \let\@oddhead\@empty
  \let\@evenhead\@empty
  \let\@oddfoot\@empty
  \let\@evenfoot\@oddfoot
}
\makeatother

\begin{document}

\global\long\def\tabfig#1{\vskip5mm \centerline{\textsc{Insert #1 around here}} \vskip5mm}


\title{Distributional uncertainty of the financial time series measured
by G-expectation\thanks{This work was supported by the National Key R\&D Program of China
(No. 2018YFA0703900) and National Natural Science Foundation of China
(No. 11701330) and Young Scholars Program of Shandong University.
Furthermore, details of this paper can be found in arXiv:2011.09226.}}

\author{\small Dedicate to P.L. CHEBYSHEV --200\\
Shige Peng \thanks{Institute of Mathematics, Shandong University, Jinan 250100, China,
(peng@sdu.edu.cn).} \quad{}Shuzhen Yang\thanks{Shandong University-Zhong Tai Securities Institute for Financial Studies,
Shandong University, PR China, (yangsz@sdu.edu.cn).} }
\date{June 24, 2021}

\maketitle

\begin{abstract}
Based on law of large numbers and central limit theorem under nonlinear expectation, we introduce a new method
of using  G-normal distribution to measure financial risks. Applying max-mean estimators and small windows method, we establish autoregressive models to determine the parameters of G-normal distribution, i.e., the return, maximal and minimal volatilities of the time series. Utilizing the value at risk (VaR) predictor model under G-normal distribution, we show that the G-VaR model gives an excellent performance in predicting the VaR for a benchmark dataset comparing to many
well-known VaR predictors.
\end{abstract}
\bigskip{}


\section{Introduction}

It is well known that many price processes in financial markets have essential
and non-negligible probability and distribution uncertainties. Here,
essential means that the such uncertainty cannot be improved. In economics,
such type of probability model uncertainty is called Knightian uncertainty,
or ambiguity. The well-known volatility uncertainty in finance markets is
a typical example.

Mathematically, such type of probability model uncertainty can be
described by a convex subset of probability measures $\{P_{\theta}\}_{\theta\in\Theta}$
defined on a measurable space $(\Omega,\mathcal{F})$ such that each
$P_{\theta}$ can possibly be a ``true” probability; however, we are
unable to distinguish the real one, and it is quite common that the
set of parameters is infinite dimensional. This means that one cannot
exhaustedly calculate and analyze each $P_{\theta}$ and subsequently find
a solution.

One solution is to introduce the notion of nonlinear expectation,
especially, sublinear expectation, and use this novel mathematical framework to address this
challenging problem. Indeed, just as indicated in the representation
theorem (see Theorem \ref{the:rep-sub}), a regular sublinear expectation is directly and
equivalently linked to a situation of convex subset of possible probability
measure $\{P_{\theta}\}_{\theta\in\Theta}$. Moreover, under a suitable
sublinear expectation space, the corresponding notion of i.i.d. and
the related law of large numbers, central limit theorem, nonlinear
normal (G-normal) distribution, maximal distributions, and the corresponding
G-Brownian motion are all naturally defined in the framework of probability
measure uncertainty.

It is easy to prove the following i.i.d. universality theorem (\citet{Peng2017,Peng2019}):

Let $\{\xi_{i}\}_{i=1}^{\infty}$ be a $d$-dimensional and uniformly
bounded random sequence in a regular sublinear expectation space $(\Omega,\mathcal{H},\mathbb{E})$.
Then, there exists a sublinear expectation $\hat{\mathbb{E}}$ on $(\Omega,\mathcal{H})$
which is stronger than $\mathbb{E}$ in the sense of
\[
\hat{\mathbb{E}}[X]\geq\mathbb{E}[X],\qquad\forall X\in H
\]
such that, under $\hat{\mathbb{E}}[\cdot]$, $\{\xi_{i}\}_{i=1}^{\infty}$
is an i.i.d. random sequence.

This theorem provides us an important flexibility of assuming the i.i.d. condition
for $\{\xi_{i}\}_{i=1}^{\infty}$ for many important problems related to
robust pricing and risk measuring.

How to estimate the sublinear distribution of a random variable $\xi$
from a given i.i.d. random sequence $\{\xi_{i}\}_{i=1}^{\infty}$,
which is a sample sequence of $\xi$? This is a very challenging and
practically important problem. Our reasoning is as follows:
It is really hard to simulate. In general, a simple
nonlinear i.i.d. sequence will become very complicated after only
few steps.
Conversely, the (sublinear) G-expectation framework is very friendly
with real life data as its sample sequence. This is in
contrast with the classical framework in which the Knightian uncertainty
is essentially neglected.

In this paper, we first give a brief presentation of the state of
arts of the theoretical framework of $G$-expectation, and, thus, provide
a theoretical argument that the log-daily
return $Z$ of NASDAQ, S\&P500 and many other bench mark stocks index obeys the law of
 $G$-normally distribution, that is, $Z\overset{d}{=}N(\mu,[\underbar{\ensuremath{\sigma}}^{2},\bar{\sigma}^{2}])$,
instead of the classical $N(\mu,  \sigma^2)$.
This means that for each function $\varphi$, the nonlinear expected
value $\mathbb{E}[\varphi(Z)]$ is a solution a fully nonlinear parabolic
PDE (G-heat equation), in which the only three parameters are $\mu,$
$\underbar{\ensuremath{\sigma}}^{2}$ and $\bar{\sigma}^{2}$.

Furthermore,
for calculating the robust VaR, we only need to compute the case $\varphi(x)=1_{[0,\infty)}(x)$
and, in this case, $\mathbb{E}[\varphi(Z)]$ has an explicit expression,
depending explicitly $\mu,$ $\underbar{\ensuremath{\sigma}}^{2}$,
and $\bar{\sigma}^{2}$.

Now, what remains is to estimate the parameters $\mu,$ $\underbar{\ensuremath{\sigma}}^{2}$
and $\bar{\sigma}^{2}$ with
\[
\mu=\mathbb{E}[Z]=-\mathbb{E}[-Z],\quad\underbar{\ensuremath{\sigma}}^{2}=-\mathbb{E}[-(Z-\mu)^{2}],\quad\bar{\sigma}^{2}=\mathbb{E}[(Z-\mu)^{2}].
\]
Using historical data of $\{Z_{t-i}\}_{i\geq1}$ and applying the
classical mean algorithm, we obtain an estimator for $\mu$; however, we
need to apply max--mean (respectively min--mean) algorithm to get estimators
for $\underbar{\ensuremath{\sigma}}^{2}$ (respectively for $\bar{\sigma}^{2}$).

The mean and variance are two important characteristics of the time
series. By allowing the volatility of a stock to range between two
extreme values $\mathrm{\sigma}_{min}$ and $\mathrm{\sigma}_{max}$,
a model of pricing and hedging derivative securities and option portfolios
was developed in \citep{ALP95}. \citep{L95} introduced optimal and
risk-free strategies for intermediaries to meet their obligations
under volatility uncertainty which takes value in some convex region
that varies with the price process, (see also \citep{DM06}, and references
therein). \citep{P97} investigated a nonlinear expectation called
$g$-expectation, which is a new formal mathematical approach to model
mean uncertainty. Continuously, \citep{CE02} used the $g$-expectation
introduced in \citep{P97} to describe the continuous-time inter-temporal
version of multiple-priors utility. Furthermore, a separate premium
for ambiguity on top of the traditional premium for risk was developed
in \citep{CE02}.

The notion of upper expectations was first discussed by \citep{Hu81}
in robust statistics, (see also \citep{Wa91}). Focusing on the measurement
of both market risks and nonmarket risks, the concept of coherent
risk measures was introduced in \citep{A99}, see also \citep{FS11}.
From a mathematical point of view, a general notation of nonlinear
expectation was originally introduced in \citep{Peng2004,Peng2005}.
A kind of nonlinear Brownian motion and related stochastic calculus
under G-expectation (a nonlinear expectation) was developed in \citep{Peng2006,Peng2008}.
A quantitative framework for defining model uncertainty in option
pricing models was considered in \citep{C06}. Through examples, the
difference between model uncertainty and the more common notion of
market risk was illustrated. \citep{KMS10}
proposed a procedure to take model risk into account in the computation
of capital reserves, which can be used to address VaR and expected
shortfall, and thus distinguish estimation risk and miss-specification
risk. Furthermore, a model of utility in a continuous-time framework
that captures aversion to ambiguity about both the volatility and
the mean of returns was formulated in \citep{EJ13}. Recently, the
systematic theory of G-expectation and a useful sublinear expectation
were concluded in monograph \citep{Peng2019}.

This paper is organized as follows:  We introduce
and explain the concept of G-expectation, nonlinear law of large numbers and central limit theorem
in Section \ref{sec:se}. In Section \ref{sec:au}, we show the $\phi$ max-mean estimation method and reveal how to implement the 1-order autoregressive models for the maximum and minimum volatilities, and
develop a Small-windows method to calculate the sample variance. In
Section \ref{sec:ea}, we use the autoregressive model to analyze the benchmark S\&P500
Index dataset, and predict the VaR by $\text{G-VaR}$ model. Finally,
we conclude this paper in Section \ref{sec:con}.

\section{Sublinear Expectation}

\label{sec:se}

Let $\Omega$ be a set and $\mathcal{H}$
be the linear space real-valued functions defined on $\Omega$ such
that, if $\xi_{1},\cdots,\xi_{d}$ in $\mathcal{H}$. Then, we also
have $\varphi(\xi_{1},\cdots,\xi_{d})\in\mathcal{H}$ for each Lipschitz
function on $\mathbb{R}^{d}$ (in many cases, we also denote by $\xi=(\xi_{1},\cdots,\xi_{d})\in\mathcal{H}^{d}$
for a $d$-dimensional random vector $\xi$). We see that $\mathcal{H}$
can be very general. It is in fact the space of our random variables.
A nonlinear expectation is functional $\mathbb{E}[\cdot]:\mathcal{H}\mapsto\mathbb{R}$
satisfies, for any given $X,Y\in\mathcal{H}$,

(i).\ \ \ $\mathbb{E}[X]\leq\mathbb{E}[Y],\ \text{if \, }X(\omega)\leq Y(\omega)$,$\quad$
for all $\omega$;

(ii). \ $\mathbb{E}[c]=c,\ c\in\mathbb{R}$.

This nonlinear expectation is called sublinear if

(iii). $\mathbb{E}[X+Y]\leq\mathbb{E}[X]+\mathbb{E}[Y]$;

(iv). $\mathbb{E}[\lambda X]=\lambda\mathbb{E}[X],\ \lambda\geq0$.

We call $\mathbb{E}$ is regular, if $\mathbb{E}[X_{i}]\downarrow0,$ for
each random variable sequence $\{X_{i}\}_{i=0}^{\infty}$, such that
$X_{i}(\omega)\downarrow0$, for all $\omega$.

We have the following generalized Daniell-Stone theorem. For further
details see Page 6, Theorem 1.2.1 of \citep{Peng2019}.

\begin{theorem}[\cite{ Peng2019}] \label{the:rep-sub} Let $\mathbb{E}[\cdot]$
be a regular sublinear expectation on $(\Omega,\mathcal{H}$). Then, there exists
a convex subset of probability measures $\{P_{\theta}\}_{\theta\in\Theta}$
defined on $(\Omega,\sigma(\mathcal{H})$), such that, for each $X\in\mathcal{H}$,
\begin{equation}
\mathbb{E}[X]=\max_{\theta\in\Theta}\int_{\Omega}X(\omega)dP_{\theta}=\max_{\theta\in\Theta}E_{P_{\theta}}[X].\label{eq:repre}
\end{equation}
\end{theorem}

This theorem indicates that a sublinear expectation can be used to
characterize the family of probabilities $\{P_{\theta}\}_{\theta\in\Theta}$,
for which a risk regulator is unable to distinguish which $P_{\theta}$
is the ``true probability.” It is quite possible that each regulator has his
own $\{P_{\theta}\}_{\theta\in\Theta}$, which is different from others.
Correspondingly, there are many different sublinear expectations.

\begin{definition}[\cite{Peng2019}] Let $\xi,\eta\in\mathcal{H}^{d}$
be two $d$-dimensional random vectors in sublinear expectation (SLE) space
$(\Omega,\mathcal{H},\mathbb{E}[\cdot])$. They are called identically
distributed, denoted by $\xi\overset{d}{=}\eta$, if
\[
\mathbb{E}[\phi(\xi)]=\mathbb{E}[\phi(\eta)],\quad\forall\phi(\cdot)\in C_{Lip}(\mathbb{R}^{d}),
\]
where $C_{Lip}(\mathbb{R}^{d})$ is the space of all Lipschitz continuous
functions on $\mathbb{R}^{d}$. We say that $\xi$ is stronger than
$\eta$ in distribution, denoted by $\xi\overset{d}{\geq}\eta$, if
\[
\mathbb{E}[\phi(\xi)]\geq\mathbb{E}[\phi(\eta)],\quad\forall\phi(\cdot)\in C_{l.Lip}(\mathbb{R}).
\]

\end{definition}

\begin{definition}[\citep{Peng2019}] A random variable $\eta\in\mathcal{H}^{n}$
is said to be independent of $\xi\in\mathcal{H}^{m}$, if for each
$\phi\in C_{Lip}(\mathbb{R}^{m}\times\mathbb{R}^{n})$, we have
\[
\mathbb{E}[\phi(\xi,\eta)]=\mathbb{E}[\mathbb{E}[\phi(x,\eta)]_{x=\xi}].
\]
\end{definition}.

\begin{definition}[\citep{Peng2019}] In a SLE
space $(\Omega,\mathcal{H},\mathbb{E})$, a sequence of $d$-dimensional
random vectors $\{\eta_{i}\}_{i=1}^{\infty}$, $\eta_{i}\in\mathcal{H}^{d}$
is said to be an i.i.d. sequence, if for each $n=1,2,\cdots$, $\eta_{n+1}$
is independent of $(\eta_{1},\cdots,\eta_{n})$. Moreover, $\eta_{1}$
and $\eta_{i+1}$ are identically distributed. \end{definition}

One can check that, if $\mathbb{E}$ is a linear expectation, then
an i.i.d. sequence under $\mathbb{E}$ is just an i.i.d. in the classical
situation under the corresponding probability measure.

A risk regulator can strengthen this subset of possible probabilities
$\{P_{\theta}\}_{\theta\in\Theta}$ in order to make a real time sequence
from real world data to be i.i.d. This is what we call the universality
of nonlinear i.i.d..

\subsection{Law of large numbers and central limit theorem}

\label{sub:gd}

Classical law of large numbers and central limit theorem play a fundamental
role in probability theory, statistic, and data analysis. Chebyshev gave a big contribution in the field of large and the related theory, i.e., Chebyshev's law of large numbers. Also see the weak law of large numbers, Bernoulli's law of large numbers, Kolmogrov's strong law of large numbers. The standard version of the central limit theorem, first given by Laplace, also see Levy-Lindeberg central limit theorem and De Moivre-Laplace central limit theorem, and comprehensive version Chebyshev's central limit theorem.

 We first introduce the maximum distribution under G-expectation.

\begin{definition}[\citep{Peng2017,Peng2019}]
A random variable $X$ under a SLE space $(\Omega,\mathcal{H},{\mathbb{E}})$
is said to be maximally distributed if there exists an interval $[\underline{\mu},\overline{\mu}]$
such that
\[
{\mathbb{E}}[\varphi(X)]=\max_{x\in[\underline{\mu},\overline{\mu}]}\varphi(x),\quad\varphi\in C_{Lip}(\mathbb{R}).
\]
We denote $X\overset{d}{=}M_{[\underline{\mu},\overline{\mu}]}$.
\end{definition}

\begin{theorem}[Law of large numbers \cite{Peng2017, Peng2019}] \label{thm17}
Let $\{X_{i}\}_{i=1}^{\infty}$ be a real valued i.i.d sequence in
a SLE space $(\Omega,\mathcal{H},{\mathbb{E}})$, satisfying
\begin{equation}
\lim_{\lambda\to+\infty}{\mathbb{E}}[(|X_{1}|-\lambda)^{+}]=0.\label{cLLN}
\end{equation}
Then, the sequence $\{\sum_{i=1}^{n}\frac{X_{i}}{n}\}_{n=1}^{\infty}$
converges to the following maximal distribution:
\begin{equation}
\lim_{n\rightarrow\infty}{\mathbb{E}}\bigg[\varphi\bigg(\sum_{i=1}^{n}\frac{X_{i}}{n}\bigg)\bigg]=\max_{\mu\in[\underline{\mu},\overline{\mu}]}[\varphi(\mu)],\quad\varphi\in C_{Lip}(\mathbb{R}),\label{eq2.13}
\end{equation}
where $\overline{\mu}={\mathbb{E}}[X_{1}],\ \underline{\mu}=-{\mathbb{E}}[-X_{1}].$
\end{theorem}
\begin{remark}
The rate of convergence of LLN and CLT under sublinear expectation plays an important role in the statistical analysis for random data under Knightian uncertainty. We refer to \citep{FPSS65}, \citep{So168}, \citep{So169}, in which an interesting nonlinear generalization of Stein method, see \cite{HPS85}, is applied as a sharp tool to attack this problem. Another important contribution to the convergence rate of G-CLT is \citep{Kr108}.  We refer to the references given in \citep{Peng2019} for other related research results.
\end{remark}

An important advantage of this nonlinear law of large numbers is that
it can be used as basis to estimate the $\mathbb{E}[\varphi(X)]$, for
each $\varphi\in C_{Lip}(\mathbb{R})$, though a nonlinear sample
sequence $\{X_{i}\}_{i=1}^{\infty}$. From the universality assumption
of i.i.d., this means that, in principle, one can obtain the nonlinear
distribution of a random variable $X$ once its sample is given.

It is clear that $\{Y_{i}\}_{i=1}^{\infty}:=\{\varphi(X_{i})\}_{i=1}^{\infty}$
is also an i.i.d. sequence in the same sublinear expectation space
$(\Omega,\mathcal{H},\mathbb{E})$. We then can take a sufficiently
large $N$ and take the mean
\[
Z_{k}:=\frac{1}{N}\sum_{i=1}^{N}Y_{Nk+i},\quad k=0,1,2,\cdots,m.
\]
It is clear that the sequence $\{Z_{k}\}_{k=0}^{m}$ is still i.i.d.
Moreover, by the above law of large numbers, for a sufficiently large
$N$, the distribution of each $Z_{k}$ converges in law to the maximal
distribution $M_{[\underline{\kappa},\overline{\kappa}]}$
with
\[
\overline{\kappa}=\mathbb{E}[Y_{1}]=\mathbb{E}[\varphi(X_{1})],\qquad\underline{\kappa}=-\mathbb{E}[-\varphi(X_{1})].
\]
Theorem \ref{cLLN}, \citep{JP16} developed the unbiased estimators
for $(\underline{\mu},\overline{\mu})$. It is then, based on the following
theorem, that we adapt the following estimator of $\mathbb{E}[\varphi(X_{1})]$,
called $\varphi$-max--mean estimator:
\[
\hat{\overline{\kappa}}=\max_{0\leq k\leq m}Z_{k}.
\]

\begin{theorem}[\cite{JP16}] \label{thm2.2} Let $\{X_{1},X_{2},\ldots,X_{n}\}$
be the $\mathbb{R}$ valued i.i.d sequence under SLE space $(\Omega,\mathcal{H},{\mathbb{E}})$,
satisfying
\[
X_{i}\overset{d}{=}M_{[\underline{\mu},\overline{\mu}]},\ i=1,\ldots,n,
\]
where $\underline{\mu}\leq\overline{\mu}$ are two unknown parameters.
Then, q.s. (quasi-surely, or $P_{\theta}$ for every $\theta\in\Theta$-almost
surely),
\[
\underline{\mu}\leq\min\{X_{1}(\omega),\ldots,X_{n}(\omega)\}\leq\max\{X_{1}(\omega),\ldots,X_{n}(\omega)\}\leq\overline{\mu},
\]
and
\[
\hat{\overline{\mu}}_{n}=\max\{X_{1},\ldots,X_{n}\}
\]
is the maximum unbiased estimate of $\overline{\mu}$,
\[
\hat{\underline{\mu}}_{n}=\min\{X_{1},\ldots,X_{n}\}
\]
is the minimum unbiased estimate of $\underline{\mu}$. \end{theorem}

Daily returns are mainly assumed to follow normal distribution. The
main theoretical support for this assumption is obviously the classical
central limit theorem: the cumulation of i.i.d. of small unknown i.i.d.
factors of $n^{-1/2}$-order will lead a normal distribution. This
argument sounds quite reasonable. But the independence assumption is obviously
too strong for prices in financial markets. In fact, it is untrue
that the distributions of the tomorrow's prices are independent of
the realization of today's prices.

Proposition 2.2.10 of \citep{Peng2019} showed that $u^{\varphi}(t,x)=\mathbb{E}[\phi(\xi_{t}+x)]$
is the unique (viscosity solution) of the following partial differential
equation:
\begin{equation}
\partial_{t}u^{\varphi}(t,x)-G(\partial_{xx}^{2}u^{\varphi}(t,x))=0,\ t>0,\ x\in\mathbb{R},\label{eq:pde-1}
\end{equation}
with the initial condition $u^{\varphi}(0,x)=\phi(x),\ x\in\mathbb{R}$,
where the function $G(\cdot)$ is defined as
\begin{equation}
G(a)=\frac{1}{2}\left(\overline{\sigma}^{2}a^{+}-\underline{\sigma}^{2}a^{-}\right),\quad a^{+}=\max(a,0),\ \text{and}\ a^{-}=\max(-a,0).\label{eq:G-function}
\end{equation}

\begin{theorem}[Central limit theorem \cite{Peng2017, Peng2019}]
\label{cCLT} Let $\{X_{i}\}_{i=1}^{\infty}$ be an $\mathbb{R}$-valued
i.i.d sequence under a SLE space $(\Omega,\mathcal{H},{\mathbb{E}})$.
We assume ${\mathbb{E}}[X_{1}]=-{\mathbb{E}}[-X_{1}]=\mu$, $\mathbb{E}[X_{1}^{2}]=\overline{\sigma}^{2}$,
$-\mathbb{E}[-X_{1}^{2}]=\underline{\sigma}^{2}$, and
\[
\lim_{\lambda\to+\infty}\mathbb{E}[(X_{1}^{2}-\lambda)^{+}]=0.
\]

Denote ${S}_{n}=\sum_{i=1}^{n}\frac{X_{i}-\mu}{\sqrt{n}}$. Then,
we have
\begin{equation}
\lim_{n\rightarrow\infty}{\mathbb{E}}[\varphi({S}_{n})]=N_{G}[\varphi]=u^{\varphi}(1,0),\quad\forall \phi\in C_{Lip}(\mathbb{R}),\label{eq2.13}
\end{equation}
where the corresponding sublinear function $G:\mathbb{R}\mapsto\mathbb{R}$
is given by:
\[
G(a):={\mathbb{E}}\bigg[\frac{1}{2}aX_{1}^{2}\bigg]=\frac{1}{2}a^{+}\overline{\sigma}^{2}-\frac{1}{2}a^{-}\underline{\sigma}^{2},\ a\in\mathbb{R}.
\]
\end{theorem}

\subsection{Robust VaR under G-normal distribution}

\label{sub:var}

We first recall the classical risk measure VaR. Let $\xi$ be a risky
position, which is a random variable probability space $(\Omega,\mathcal{F},P)$,
where the probability measure $P$ is given then the VaR of $\xi$,
for given $\alpha\in(0,1)$, denoted by VaR$_{\alpha}(\xi)$, is defined
as
\[
\text{VaR}_{\alpha}(\xi):=-\inf\{x\in\mathbb{R}:E_{P}[1_{\xi\leq x}]>\alpha\}.
\]
It is clear that the value of $\text{VaR}{}_{\alpha}(\xi)$ depends
on the given probability distribution of $\xi$.

If this risky position $\xi$ is $G$-normally distributed, say, $\xi\overset{d}{=}N(\mu,[\underline{\sigma}^{2},\overline{\sigma}^{2}])$,
then the robust $G$-VaR is:
\begin{align*}
G-\text{VaR}_{\alpha}(\xi): & =-\inf\{x\in\mathbb{R}:\max_{\theta\in\Theta}E_{P_{\theta}}[\mathbf{1}_{\xi\leq x}]>\alpha\}\\
 & =-\inf\{x\in\mathbb{R}:\mathbb{E}_{G}[\mathbf{1}_{\xi\leq x}]>\alpha\}\\
 & =-\inf\{x\in\mathbb{R}:\hat{F}_{G}(x)>\alpha\},
\end{align*}
where $\hat{F}_G(x)$ is:
\[
\hat{F}_{G}(x)=\mathbb{E}_{G}[\mathbf{1}_{\xi\leq x}]=u(1,x)
\]
and $u(t,x)$ is the solution of the following
\[
\partial_{t}u(t,x)=G(\partial_{xx}^{2}u(t,x)),\quad u(0,x)=\mathbf{1}_{[0,\infty)}(x),\quad t\in(0,1],\;x\in\mathbb{R}.
\]
It is practically important that $u(1,x)$ has the following explicit
expression:
\begin{equation}
\hat{F}_G(x)=\frac{2\overline{\sigma}}{\overline{\sigma}+\underline{\sigma}}\Phi\left(\frac{x-\mu}{\overline{\sigma}}\right)\ \mathbf{1}_{(-\infty,0]}(x)+\left\{ 1-\frac{2\bar{\sigma}}{\overline{\sigma}+\underline{\sigma}}\Phi\left(-\frac{x-\mu}{\underline{\sigma}}\right)\right\} \ \mathbf{1}_{(0,\infty)}(x),\label{eq:Fhat20}
\end{equation}
where $\Phi(\cdot)$ is the distribution function of the standard
normal distribution. The G-VaR model is
\begin{equation}
\text{G-VaR}_{\alpha}(\xi)=-\hat{F}_G^{-1}(\alpha),\label{eq:GVaR2}
\end{equation}
which can be directly and simply calculated for $\xi\overset{d}{=}N(\mu,[\underbar{\ensuremath{\sigma}}^{2},\bar{\sigma}^{2}])$.
Observe that this G-VaR value depends only on $\alpha$ and the parameters
$\mu,\underbar{\ensuremath{\sigma}},\bar{\sigma}.$ We also denote
\[
\text{G-VaR}_{\alpha}[\mu,\underbar{\ensuremath{\sigma}},\bar{\sigma}]=\text{G-VaR}_{\alpha}(\xi).
\]

\section{Estimating $\mu$, $\underbar{\ensuremath{\sigma}}$, $\bar{\sigma}$
for G-normally distributed i.i.d. samples}
\label{sec:au}

\subsection{A robust max-mean estimators }

Let $\{Z_{t-i}\}_{1\leq i\leq n}$ be the historical data before time $t$,
which is assumed to be i.i.d. with $Z_{t}\overset{d}{=}N(\mu,[\underbar{\ensuremath{\sigma}}^{2},\bar{\sigma}^{2}])$
in a SLE space $(\Omega,\mathcal{H},\mathbb{E})$.
We show how to estimate the parameters $r$, $\bar{\sigma}^{2}$
and $\underbar{\ensuremath{\sigma}}^{2}$. In fact, following the
max--mean rule, for the mean value of $Z_t$, which is
\[
\mu=\mathbb{E}[Z_{t}]=-\mathbb{E}[-Z_{t}],
\]
we just need an average for each window of sample, namely
\[
\hat{r}_{t,j}=\frac{\sum_{i=1}^{L}Z_{t-L+i-j}}{L}.
\]
However, to estimate
\[
,\qquad\bar{\sigma}^{2}=\mathbb{E}[(Z_{1}-\mu)^{2}],\quad\underbar{\ensuremath{\sigma}}^{2}=-\mathbb{E}[-(Z_{1}-\mu)^{2}].
\]
We need first calculate the local variance within the same window, that is,
\[
\begin{array}{l}
\tilde{\sigma}_{t,j}^{2}={\displaystyle \frac{\sum_{i=1}^{L}(Z_{t-L+i-j}-\hat{r}_{t,j})^{2}}{L-1}}.\end{array}
\]
Then estimators of upper and lower variances are defined by:
\[
\hat{\underline{\sigma}}_{t}^{2}=\min_{0\leq j\leq K-1}\tilde{\sigma}_{t,j}^{2}\quad\quad\text{and}\quad\quad\hat{\overline{\sigma}}_{t}^{2}=\max_{0\leq j\leq K-1}\tilde{\sigma}_{t,j}.
\]
An important problem is how to choose the window's width $L$ and
number $K$. According to our analysis in the previous section, we
need to take a sufficiently big $L$. In fact, this type of estimators
had been successfully applied to the corresponding $G$-VaR algorithm.
We then use this algorithm for the case where the data $\{Z_{i}\}_{i=-N}^{t}$ is the
log-daily returns of NASDAQ. The performance of this $G$-VaR model
was more than excellent comparing to many existing high-ranking VaR
algorithms(see \cite{Peng2020} for details).

\subsection{The case for Small-windows }

We have also investigated the case of small size $K$ and small window number $L$ applied
estimated parameters to get the corresponding $G$-VaR value, using the stock index S\&P500 data.
The results are not so good. But then we tried to apply a very simple 1st order autoregressive model,
separately to the sequences $\{\hat{r}_{t,K-1}\}$,  $\{\hat{\underline{\sigma}}_{t}^{2}\}$, and $\{\hat{\overline{\sigma}}_{t}^{2}\}$, namely, we apply the classical least mean-square approach to
\begin{equation}
{r}_{t}={\gamma}_{0}+\gamma_{1}r_{t-1}+\varepsilon_{1,t},\label{eq:kap11-1}
\end{equation}
\begin{equation}
\overline{\sigma}_{t}^{2}={\alpha}_{0}+{\alpha}_{1}\overline{\sigma}_{t-1}^{2}+\varepsilon_{2,t},\label{eq:sig10-1}
\end{equation}
\begin{equation}
{\underline{\sigma}}_{t}^{2}={\beta}_{0}+\beta_{1}\underline{\sigma}_{t-1}^{2}++\varepsilon_{3,t},\label{eq:kap10-1}
\end{equation}
and use the 1-step predicted $({r}_{t}, \overline{\sigma}_{t}^{2}, {\underline{\sigma}}_{t}^{2})$. Then, we
get surprisingly excellent results for the case of log-daily return of stock index S\& P500 data, with window with $K$ and windows number $L$, in
\[
K\in \{5,6,\cdots,15\}\,\,\, \text{and}\,\,\, L\in  \{5,6,\cdots,15\}.
\]

For reader's convenience, we conclude the algorithm of G-VaR of a time series $\{Z_{s}\}$ at
time $t+1$ as follows \footnote{The MATLAB code of the algorithms can be found in https://github.com/yangsz-code/G-VaRII.
There are many advantages of our method G-VaR: $(i)$, it is easy
to achieve the algorithms of G-VaR; $(ii)$, we only need to estimate
three parameters by classical least square estimate; $(iii)$, the
algorithms of G-VaR are very efficient.}:
\medskip
\medskip

\begin{algorithm}[H]
\caption{The algorithms of G-VaR}
\textbf{Step 1:} Download the stock time sequence data, for example,
the value of daily data of stock index S\&P500 Index;

\textbf{Step 2:} Calculate the log-returns daily data of the stock,
and denotes as $\{Z_{s}\}_{s\leq t}$. Based on the small-windows
$(K,L)\in A\times B$, for a given risk level $\alpha$, we can find
$(K,L)$, and use data $\{Z_{s-i}\}_{i=0}^{L+K-2},\ t-N+1\leq s\leq t$
to calculate the estimators $\{(\hat{r}_{s},\hat{\overline{\sigma}}_{s}^{2},\hat{\underline{\sigma}}_{s}^{2})\}_{s=t-N+1}^{t}$;

\textbf{Step 3:} Then, based on the estimators $\{(\hat{r}_{s},\hat{\overline{\sigma}}_{s}^{2},\hat{\underline{\sigma}}_{s}^{2})\}_{s=t-N+1}^{t}$,
and first-order autoregressive modes (\ref{eq:kap11-1}), (\ref{eq:sig10-1}) and (\ref{eq:kap10-1}), applying least square estimator method, we
can obtain the coefficients $(\hat{\gamma},\hat{\alpha},\hat{\beta})$;

\textbf{Step 4:} The one-period forecast mean, maximum variance and
minimum variance at time $t+1$ are given by the autoregressive models. By $(\tilde{r}_{t+1},\tilde{\overline{\sigma}}_{t+1}^{2},\tilde{\underline{\sigma}}_{t+1}^{2})$,
we can obtain the G-VaR at time $t+1$.
\end{algorithm}

\section{Empirical analysis}
\label{sec:ea}

Now, we use the 1-order autoregressive models
to study the stock index S\&P500 Index\footnote{The dataset are downloaded from https://finance.yahoo.com/lookup.}.
We denote the 100 times log-returns daily data of S\&P500 Index as
$Z$, and one observation sequence is from
January 4, 2010 to July 17, 2020, with a total of $n=2653$ daily
data. $Z_{t}$ obeys
the G-normal distribution $N(r_{t},[\underline{\sigma}_{t}^{2},\overline{\sigma}_{t}^{2}])$.
The return $r_{t}$, maximum volatility $\overline{\sigma}_{t}$,
and minimum volatility $\underline{\sigma}_{t}$ satisfy the 1-order autoregressive modes (\ref{eq:kap11-1}), (\ref{eq:sig10-1}) and (\ref{eq:kap10-1}).

For the given $K$ and $L$,
we assume the samples $\{Z_{s-i}\}_{i=0}^{L+K-2}$ are independent
under sublinear expectation $\mathbb{E}[\cdot]$, and we can use data $\{Z_{s-i}\}_{i=0}^{L+K-2}$ to obtain a sample
estimator $\hat{\overline{\sigma}}_{s}$ for the maximum volatility
$\overline{\sigma}_{s}$, estimator $\hat{\underline{\sigma}}_{s}$
for volatility uncertainty index $\underline{\sigma}_{s}$, and the
data $\{Z_{s-i}\}_{i=0}^{L-1}$ to obtain a sample estimator $\hat{r}_{s}$
for return $r_{s}$. Based on the least square estimator and sample
estimators $\{\hat{\overline{\sigma}}_{s},\hat{\underline{\sigma}}_{s},\hat{r}_{s}\}_{s=t-N+1}^{t}$,
we can obtain the estimators $(\hat{{\alpha}}_{0},\hat{{\alpha}}_{1})$,
$(\hat{\beta}_{0},\hat{\beta}_{1})$ and $(\hat{\gamma}_{0},\hat{\gamma}_{1})$
for coefficients $({\overline{\alpha}}_{0},{\overline{\alpha}}_{1})$ via least square estimator,
$({\beta}_{0},{\beta}_{1})$ and $({\gamma}_{0},{\gamma}_{1})$, and
one-period forecast estimators $(\tilde{\overline{\sigma}}_{t+1},\tilde{\underline{\sigma}}_{t+1},\tilde{r}_{t+1})$
for $(\overline{\sigma}_{t+1},\underline{\sigma}_{t+1},r_{t+1})$. When the samples $\{Z_{t-i}\}_{i=0}^{L+K-2}$ are i.i.d. under the sublinear expectation $\mathbb{E}[\cdot]$,
based on Theorem \ref{thm2.2}, we can obtain that $\hat{\overline{\sigma}}_{t}$
is an unbiased estimator for the maximum volatility $\overline{\sigma}_{t}$,
and $\hat{\underline{\sigma}}_{t}$ is an unbiased estimator for the
minimum volatility $\underline{\sigma}_{t}$\footnote{For the statistical method under sublinear expectation $\mathbb{E}[\cdot]$,
see \citep{JP16}. This estimation method is called Max--Mean
calculation in \citep{Peng2017}; also see Section 5 of \citep{Peng2020}.}.

In the following, we take $K=5,L=10,N=100$ and verify the autoregressive
properties of $\{\hat{\overline{\sigma}}_{s},\hat{\underline{\sigma}}_{s},\hat{r}_{s}\}_{s=t-N+1}^{t}$.
Based on the least square estimator, we
can obtain the coefficients, $(\hat{\overline{\alpha}}_{0},\hat{\overline{\alpha}}_{1})=(0.1888,0.9861)$,
$(\hat{\beta}_{0},\hat{\beta}_{1})=(0.2111,0.9750)$ and $(\hat{\gamma}_{0},\hat{\gamma}_{1})=(0.0049,0.8373)$.
Hence, the one-period forecast parameters $(\tilde{\overline{\sigma}}_{t+1},\tilde{\underline{\sigma}}_{t+1},\tilde{r}_{t+1})$
is given as follows:
\[
\begin{array}{ll}
 & \tilde{\overline{\sigma}}_{t+1}^{2}=0.1888+0.9861\hat{\overline{\sigma}}_{t}^{2},\\
 & \tilde{{\underline{\sigma}}}_{t+1}^{2}=0.2111+0.9750\hat{\underline{\sigma}}_{t}^{2},\\
 & \tilde{r}_{t+1}=0.0049+0.8373\hat{r}_{t}.
\end{array}
\]

\begin{figure}[H]
\centering \includegraphics[width=4.5in]{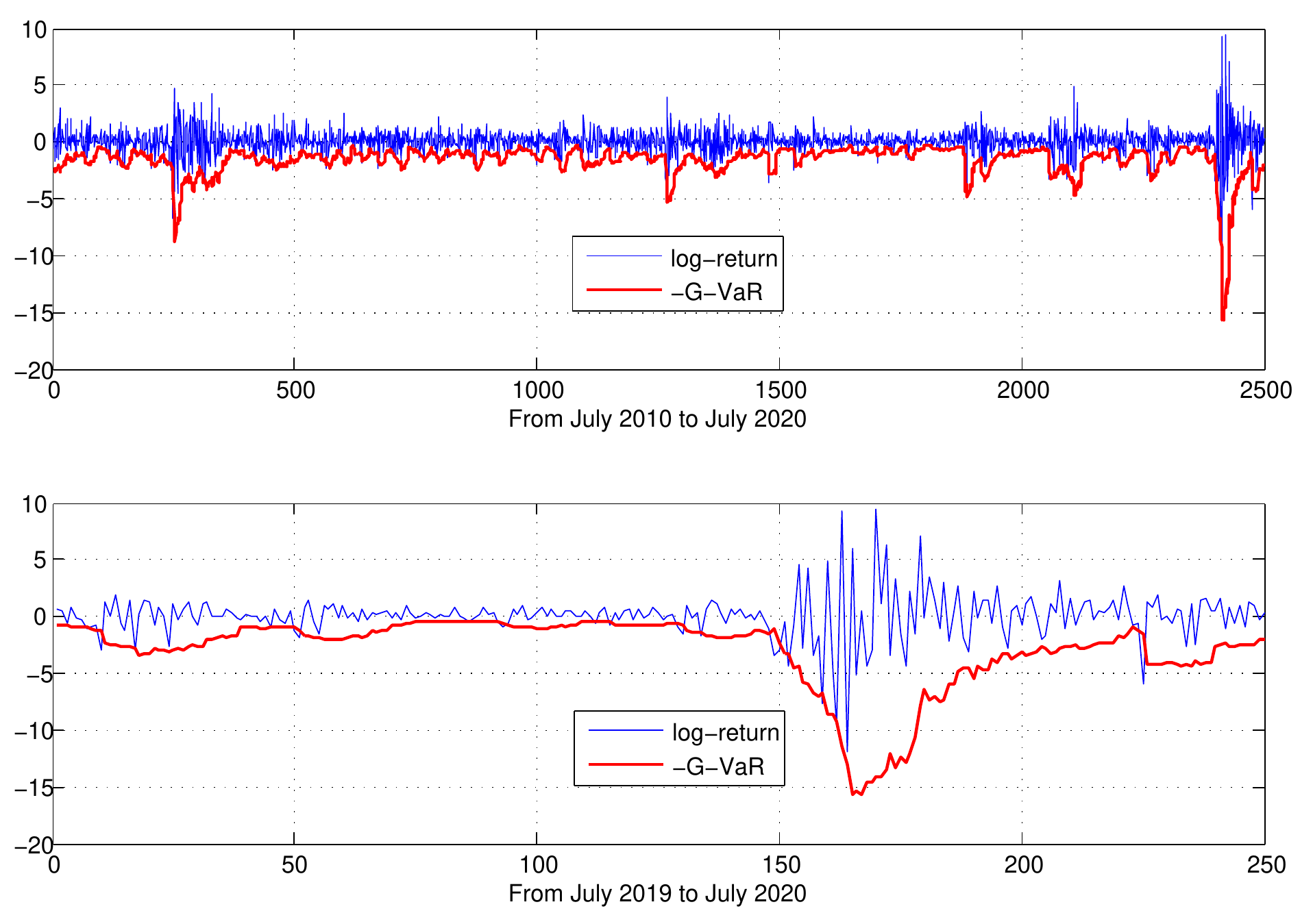} \caption{$\text{G-VaR}$ value for log-return of S\&P500 Index and the dynamic
value of log-returns of S\&P500 Index with risk level $\alpha=0.05$.
\label{fig:var1}}
\end{figure}

In the first picture of Figure \ref{fig:var1}, the blue line denotes
the log-returns of S\&P500 Index, and the red line denotes the VaR
under $-\text{G-VaR}_{\alpha}(Z_{t+1})$ model,
and the time $t+1$ is from July 2010 to July 2020. Moreover, as per
the second picture of Figure \ref{fig:var1}, the measure $\text{G-VaR}_{\alpha}(Z_{t+1})$
can capture the local changes of the log-returns of S\&P500 Index.

We consider small-windows $5\leq K,L\leq15$, which are used
to estimate the maximum volatility $\tilde{\overline{\sigma}}_{t+1}$
and the volatility uncertainty index $\tilde{\kappa}_{t+1}$. The
excellent performance of the G-VaR model in Figure \ref{fig:var1}
indicates that the worst-case distribution can capture the local changing
of a log-return of S\&P500 Index. Therefore, we comment that the small
windows $K$ and $L$ are important for estimating the local maximum
volatility and volatility uncertainty index. Furthermore, the one-period
forecast maximum volatility $\tilde{\overline{\sigma}}_{t+1}$ and
volatility uncertainty index $\tilde{\kappa}_{t+1}$ can be used to
capture the quantile of the distribution of the log-returns of S\&P500
Index. The count numbers of the violations of the log-returns of S\&P500
Index are given as follows:
\[
\begin{array}{ll}
 & m_{00}(h)=\#\{-\text{G-VaR}_{\alpha}(Z_{t+1})<Z_{t+1},-\text{G-VaR}_{\alpha}(Z_{t+2})<Z_{t+2},\ t_{0}<t\leq h\};\\
 & m_{01}(h)=\#\{-\text{G-VaR}_{\alpha}(Z_{t+1})<Z_{t+1},-\text{G-VaR}_{\alpha}(Z_{t+2})>Z_{t+2},\ t_{0}<t\leq h\};\\
 & m_{10}(h)=\#\{-\text{G-VaR}_{\alpha}(Z_{t+1})>Z_{t+1},-\text{G-VaR}_{\alpha}(Z_{t+2})<Z_{t+2},\ t_{0}<t\leq h\};\\
 & m_{11}(h)=\#\{-\text{G-VaR}_{\alpha}(Z_{t+1})>Z_{t+1},-\text{G-VaR}_{\alpha}(Z_{t+2})>Z_{t+2},\ t_{0}<t\leq h\},
\end{array}
\]
where $t_{0}$ is the initial date which is used to forecast $\text{G-VaR}_{\alpha}(Z_{t_{0}+1})$,
and $\#\{\cdot\}$ denotes the count numbers which satisfies the violation
conditions. Furthermore, we set
\[
m_{1}(h)=m_{11}(h)+m_{10}(h),\ m_{0}(h)=m_{00}(h)+m_{01}(h),
\]
and
\[
\begin{array}{ll}
 & {\displaystyle \hat{\alpha}(h)=\frac{m_{1}(h)}{m_{0}(h)+m_{1}(h)},\quad1-\hat{\alpha}(h)=\frac{m_{0}(h)}{m_{0}(h)+m_{1}(h)};}\\
 & {\displaystyle \pi_{01}(h)=\frac{m_{01}(h)}{m_{00}(h)+m_{01}(h)},\quad\pi_{11}(h)=\frac{m_{11}(h)}{m_{10}(h)+m_{11}(h)};}\\
 & {\displaystyle \pi(h)=\frac{m_{01}(h)+m_{11}(h)}{m_{00}(h)+m_{01}(h)+m_{10}(h)+m_{11}(h)}.}
\end{array}
\]
To assess the predictive performance of the $\text{G-VaR}(Z_{t+1})$
model, we use the test of a likelihood ratio for a Bernoulli trial
and the test of a Christofferson independent to verify it. Let $\hat{\alpha}(h)$
be the sample violations rate and denote the likelihood ratio test
statistics,
\[
\mathbb{T}_{1}(h)=2m_{1}(h)\ln\frac{\hat{\alpha}(h)}{\alpha}+2m_{0}(h)\ln\frac{1-\hat{\alpha}(h)}{1-\alpha},
\]
and the Christofferson independent test statistics,
\[
\mathbb{T}_{2}(h)=2\ln\left[\frac{(1-\pi_{01}(h))^{m_{00}(h)}(\pi_{01}(h))^{m_{01}(h)}(1-\pi_{11}(h))^{m_{10}(h)}(\pi_{11}(h))^{m_{11}(h)}}{(1-\pi(h))^{m_{00}(h)+m_{10}(h)}(\pi(h))^{m_{01}(h)+m_{11}(h)}}\right].
\]
Applying the well-known asymptotic $\chi^{2}(1)$ distribution, the
$p$-value of the test is,
\[
\text{LR}_{uc}^{h}=P\left(\chi^{2}(1)>\mathbb{T}_{1}(h)\right),
\]
and independent test of violations point,
\[
\text{LR}_{ind}^{h}=P\left(\chi^{2}(1)>\mathbb{T}_{2}(h)\right).
\]

In the following, we conclude the testing results of S\&P500 Index
with $\alpha=0.05$. 
\begin{table}[H]
\centering \caption{Testing results of S\&P500 Index with $\alpha=0.05$}
\label{table:sp1} %
\begin{tabular}{ccccccc}
\toprule
Model  & Time  & $h-t_{0}$  & $\hat{\alpha}(h)$  & $\text{LR}_{uc}^{h}$  & $\text{LR}_{ind}^{h}$  & $100\overline{\text{VaR}}$\tabularnewline
\midrule
G-VaR:  & 201907-202007  & 250  & 0.068  & 0.215  & 0.115  & 2.987 \tabularnewline
 & 201607-202007  & 1000  & 0.048  & 0.770  & 0.102  & 1.685\tabularnewline
 & 201007-202007  & 2500  & 0.052  & 0.715  & 0.890  & 1.640\tabularnewline
\bottomrule
\end{tabular}
\end{table}

In Table \ref{table:sp1}, we consider three lengths of dates---one
year, four years, and ten years. We calculate the indices $\hat{\alpha}(h)$,
$\text{LR}_{uc}^{h}$, $\text{LR}_{ind}^{h}$, and $100\overline{\text{VaR}}$,
respectively, where $100\overline{\text{VaR}}$ is 100 times VaR under
$\text{G-VaR}$. It should be noted that
\[
\text{LR}_{uc}^{h}>0.05,\quad\text{LR}_{ind}^{h}>0.05,
\]
thus, the likelihood ratio test statistics and the Christofferson independent
test statistics are through under the confidential level $95\%$.

\subsection{Comparison with others VaR models}

\label{sec:comp} As \citep{Kuester06} pointed out, the best
VaR predictions for the benchmark are obtained by AR-GARCH filtered
modeling such as the recommended AR-GARCH-Skewed-t or AR-GARCH-Skewed-t-EVT
models. In \citep{Peng2020}, the G-VaR predictor is compared with
these two models and a more traditional AR-GARCH-Normal predictor.
Hence, we use the dataset S\&P500 Index from January 3, 2000 to February
7, 2018 in \citep{Peng2020}, to compare the models developed in this
study with the above four models. We denote ${\text{G-VaR}}^{*}$
as the VaR model developed in \citep{Peng2020}. Partial results of
Table \ref{table:sp3} were downloaded from Table 7 of \citep{Peng2020}
with historical data window $W=250$. In Table \ref{table:sp3}, the
G-VaR model with parameter $(K,L,N)=(5,10,100)$ under risk level
$\alpha=0.05$ and $(K,L,N)=(6,5,100)$ under risk level $\alpha=0.01$.
\begin{table}[H]
\centering \caption{Testing results of S\&P500 Index under VaR models}
\label{table:sp3} %
\begin{tabular}{lcccccc}
\toprule
Model  & Times  & $\alpha$ & $\hat{\alpha}(h)$  & $\text{LR}_{uc}^{h}$  & $\text{LR}_{ind}^{h}$  & $100\overline{\text{VaR}}$\tabularnewline
\midrule
$\text{G-VaR}$:  & 200101-201802  & 0.05  & 0.051  & \textbf{0.84}  & \textbf{0.99}  & 1.87\tabularnewline
 & 200101-201802  & 0.01  & 0.011  & \textbf{0.76}  & \textbf{1.00}  & 3.02\tabularnewline
${\text{G-VaR}}^{*}$:  & 200101-201802  & 0.05  & 0.050  & \textbf{0.88}  & 0.00  & 1.83\tabularnewline
 & 200101-201802  & 0.01  & 0.010  & \textbf{0.87}  & 0.00  & 3.46\tabularnewline
GARCH(1,1)-N:  & 200101-201802  & 0.05  & 0.062  & 0.00  & \textbf{0.62}  & 1.65\tabularnewline
 & 200101-201802  & 0.01  & 0.026  & 0.00  & \textbf{0.19}  & 2.35\tabularnewline
GARCH(1,1)-St:  & 200101-201802  & 0.05  & 0.060  & 0.01  & \textbf{0.74}  & 1.69\tabularnewline
 & 200101-201802  & 0.01  & 0.014  & 0.01  & \textbf{0.23}  & 2.68\tabularnewline
GARCH(1,1)-St-EVT:  & 200101-201802  & 0.05  & 0.052  & \textbf{ 0.59}  & \textbf{0.40} & 1.80\tabularnewline
 & 200101-201802  & 0.01  & 0.014  & 0.01  & \textbf{0.06}  & 2.71\tabularnewline
\bottomrule
\end{tabular}
\end{table}

In Table \ref{table:sp3}, we show the testing results of S\&P500
Index under the VaR models: $\text{G-VaR}$, ${\text{G-VaR}}^{*}$,
GARCH(1,1)-N, GARCH(1,1)-St, and GARCH(1,1)-St-EVT. We denote the
values of the likelihood ratio test $\text{LR}_{uc}^{h}$ and the
Christofferson independent test $\text{LR}_{ind}^{h}$ as boldface
Type, which are through the tests under the confidential level $95\%$.
We can see that the ${\text{G-VaR}}^{*}$ model in \citep{Peng2020}
is the best model among the above VaR models based on the value
$\text{LR}_{uc}^{h}$. ${\text{G-VaR}}^{*}$ model can capture
the long-time average loss of S\&P500 Index. In fact, it should be
noted that $\text{LR}_{uc}^{h}$ can test the accuracy of the VaR
model and $\text{LR}_{ind}^{h}$ can test the independence of the
violation points. Combining the values of $\text{LR}_{uc}^{h}$ and
$\text{LR}_{ind}^{h}$, we can see that models $\text{G-VaR}$ and
GARCH(1,1)-St-EVT are the best for $\alpha=0.05$, and $\text{G-VaR}$
is the best for $\alpha=0.01$. In particular, we also use the $\text{G-VaR}$
model to predict VaR for NASDAQ Composite Index and CSI300, and we
have a performance similar to that of $\text{G-VaR}$ model, as shown in
Table \ref{table:sp3}. Hence, we can see that the autoregressive
models developed in this study are powerful tools to investigate the
excellent performance of the $\text{G-VaR}$ model.

\section{Conclusion}

\label{sec:con}


Sublinear expectation is a powerful theoretical framework for measuring
the probabilistic and distributional uncertainties inherent in many time sequences.
Based on the central limit theorem in the theory of sublinear expectation,
many important financial time series need to be assumed to follow the law of
$G$-normal distribution in which the volatility uncertainty is naturally
taken into account. It turns out that compared
with many widely used VaR algorithms, the corresponding $G$-VaR
is a much more robust and efficient financial risk measure (see Section \ref{sec:ea}).

%

Based on the assumption that the time series satisfies a G-normal
distribution under sublinear expectation, we use a worst-case
distribution to capture the robust distribution property of the time series,
where the G-normal distribution denotes an infinite family of distributions.
In \citep{Peng2020}, the empirical analysis is given based on an
adaptive window $W_{0}$, which can capture the long-time loss of
the time series, but is unable to follow the local changing of the time
series. Diverging from \citep{Peng2020}, we develop a concept of
Localization Windows to estimate the local sample variance. Then,
combining the autoregressive models, we can obtain a one-period forecast
maximum volatility, minimum volatility, and volatility uncertainty
index. In Section \ref{sec:ea}, we apply this new mechanism to predict
the VaR of S\&P500 Index by the $\text{G-VaR}_{\alpha}(\cdot)$ model,
which shows the advantage  of using the autoregressive models and the Localization
Windows.



\bibliography{gexp1}

\begin{thebibliography}{26}
\providecommand{\natexlab}[1]{#1}
\providecommand{\url}[1]{\texttt{#1}}
\expandafter\ifx\csname urlstyle\endcsname\relax
  \providecommand{\doi}[1]{doi: #1}\else
  \providecommand{\doi}{doi: \begingroup \urlstyle{rm}\Url}\fi

\bibitem[Artzner et~al.(1999)Artzner, Delbaen, Eber, and Heath]{A99}
P.~Artzner, F.~Delbaen, J.-M. Eber, and D.~Heath.
\newblock Coherent measures of risk.
\newblock \emph{Mathematical Finance}, 9:\penalty0 203--228, 1999.

\bibitem[Avellaneda et~al.(1995)Avellaneda, Levy, and Par\'{a}s]{ALP95}
M.~Avellaneda, A.~Levy, and A.~Par\'{a}s.
\newblock Pricing and hedging derivative securities in markets with uncertain
  volatilities.
\newblock \emph{Applied Mathematical Finance}, 2:\penalty0 73--88, 1995.

\bibitem[Chen and Epstein(2002)]{CE02}
Z.~Chen and L.~Epstein.
\newblock Ambiguity, risk, and asset returns in continuous time.
\newblock \emph{Econometrica}, 70\penalty0 (4):\penalty0 1403--1443, 2002.

\bibitem[Cont(2006)]{C06}
R.~Cont.
\newblock Model uncertainty and its impact on the pricing of derivative
  instruments.
\newblock \emph{Mathematical Finance}, 16:\penalty0 519--547, 2006.

\bibitem[Denis and Martini(2006)]{DM06}
L.~Denis and C.~Martini.
\newblock A theoretical framework for the pricing of contingent claims in the
  presence of model uncertainty.
\newblock \emph{Annal of Applied Probability}, 16:\penalty0 827--852, 2006.

\bibitem[Epstein and Ji(2013)]{EJ13}
L.~G. Epstein and S.~Ji.
\newblock Ambiguous volatility and asset pricing in continuous time.
\newblock \emph{Review of Financial Studies}, 26\penalty0 (7):\penalty0
  1740--1786, 2013.

\bibitem[Fang et~al.(2019)Fang, Peng, Shao, and Song]{FPSS65}
X.~Fang, S.G. Peng, Q.M. Shao, and Y.S. Song.
\newblock Limit theorems with rate of convergence under sublinear expectations.
\newblock \emph{Bernoulli}, 25(4A):\penalty0 2564--2596, 2019.

\bibitem[F\"{o}llmer and Schied(2011)]{FS11}
H.~F\"{o}llmer and A.~Schied.
\newblock \emph{Stochastic Finance. An Introduction in Discrete Time. Third
  revised and extended edition}.
\newblock Walter de Gruyter \& Co., Berlin, 2011.

\bibitem[Hu et~al.(2017)Hu, Peng, and Song]{HPS85}
M.S. Hu, S.G. Peng, and Y.S. Song.
\newblock Stein type characterization for g-normal distributions.
\newblock \emph{Electron. Commun. Probab.}, 22:\penalty0 1--12, 2017.

\bibitem[Huber(1981)]{Hu81}
P.~J. Huber.
\newblock \emph{Robust Statistics. Wiley Series in Probability and Mathematical
  Statistics}.
\newblock John Wiley \& Sons, Inc., New York, 3rd edition, 1981.

\bibitem[Jin and Peng(2016)]{JP16}
H.~Jin and S.~Peng.
\newblock Optimal unbiased estimation for maximal distribution.
\newblock \emph{arXiv:1611.07994v1}, 2016.

\bibitem[Kerkhof et~al.(2010)Kerkhof, Melenberg, and Schumacher]{KMS10}
J.~Kerkhof, B.~Melenberg, and H.~Schumacher.
\newblock Model risk and capital reserves.
\newblock \emph{Journal of Banking \& Finance}, 34:\penalty0 267--279, 2010.

\bibitem[Krylov(2020)]{Kr108}
N.V. Krylov.
\newblock On shige peng's central limit theorem.
\newblock \emph{Stochastic Processes and their Applications}, 130(3):\penalty0
  1426--1434, 2020.

\bibitem[Kuester et~al.(2006)Kuester, Mittnik, and Paolella]{Kuester06}
K.~Kuester, S.~Mittnik, and M.~S. Paolella.
\newblock \text{Value-at-Risk Prediction}: A comparison of alternative
  strategies.
\newblock \emph{Journal of Financial Econometrics}, 4\penalty0 (1):\penalty0
  53--89, 2006.

\bibitem[Lyons(1995)]{L95}
T.~J. Lyons.
\newblock Uncertain volatility and the risk-free synthesis of derivatives.
\newblock \emph{Applied Mathematical Finance}, 2:\penalty0 117--133, 1995.

\bibitem[Peng(1997)]{P97}
S.~Peng.
\newblock \emph{Backward SDE and related g-expectation. Backward stochastic
  differential equations (Paris, 1995-1996) 141-159}.
\newblock Pitman Res. Notes Math. Ser., Longman, Harlow, 1997.

\bibitem[Peng(2004)]{Peng2004}
S.~Peng.
\newblock Filtration consistent nonlinear expectations and evaluations of
  contingent claims.
\newblock \emph{Acta Mathematicae Applicatae Sinica}, 20:\penalty0 1--24, 2004.

\bibitem[Peng(2005)]{Peng2005}
S.~Peng.
\newblock Nonlinear expectations and nonlinear {M}arkov chains.
\newblock \emph{Acta Mathematicae Applicatae Sinica}, 26B:\penalty0 159--184,
  2005.

\bibitem[Peng(2006)]{Peng2006}
S.~Peng.
\newblock \emph{Stochastic Analysis and Applications: The Abel Symposium 2005},
  chapter "$G$-expectation, $G$-{B}rownian motion and related stochastic
  calculus of {I}t\^o type".
\newblock Springer, Berlin Heidelberg, 2006.

\bibitem[Peng(2008)]{Peng2008}
S.~Peng.
\newblock Multi-dimensional \text{G}-{B}rownian motion and related stochastic
  calculus under \text{G}-expectation.
\newblock \emph{Stochastic Processes and Their Applications}, 118:\penalty0
  2223--2253, 2008.

\bibitem[Peng(2017)]{Peng2017}
S.~Peng.
\newblock Theory, methods and meaning of nonlinear expectation theory.
\newblock \emph{Scientia Sinica Mathematica: In Chinese}, 47:\penalty0
  1223--1254, 2017.

\bibitem[Peng(2019)]{Peng2019}
S.~Peng.
\newblock \emph{Nonlinear {E}xpectations and {S}tochastic {C}alculus {u}nder
  {U}ncertainty}, pages 1--212.
\newblock Springer, Berlin, Heidelberg, 2019.

\bibitem[Peng et~al.(2020)Peng, Yang, and Yao]{Peng2020}
S.~Peng, S.~Yang, and J.~Yao.
\newblock Improving value-at-risk prediction under model uncertainty.
\newblock arXiv:1805.03890:\penalty0 1--42, 2020.

\bibitem[Song(2019)]{So169}
Y.S. Song.
\newblock Stein's method for law of large numbers under sublinear expectations.
\newblock \emph{arXiv:1904.04674}, pages 1--12, 2019.

\bibitem[Song(2020)]{So168}
Y.S. Song.
\newblock Normal approximation by stein's method under sublinear expectations.
\newblock \emph{Stochastic Processes and their Applications}, 130(5):\penalty0
  2838--2850, 2020.

\bibitem[Walley(1991)]{Wa91}
P.~Walley.
\newblock \emph{Statistical Reasoning with Imprecise Probabilities}.
\newblock Chapman and Hall, 1991.

\end{thebibliography}

\end{document}